\documentclass[preprint,showpacs,prl]{revtex4}
\usepackage{amssymb}
\usepackage[dvipdfm]{graphicx}
\usepackage{color}

\begin{document}

\title{Energetic disorder at the metal/organic semiconductor interface}
\author{S.V. Novikov}
\affiliation{A.N. Frumkin Institute of Physical Chemistry and
Electrochemistry, Moscow 119991, Russia}
\author{G.G. Malliaras}
\affiliation{Materials Science and Engineering, Cornell
University, Ithaca, NY 14853-1501}


\begin{abstract}
The physics of organic semiconductors is dominated by the effects
of energetic disorder. We show that image forces reduce the
electrostatic component of the total energetic disorder near an
interface with a metal electrode. Typically, the variance of
energetic disorder is dramatically reduced at the first few layers
of organic semiconductor molecules adjacent to the metal
electrode. Implications for charge injection into organic
semiconductors are discussed.
\end{abstract}

\maketitle

\textcolor{red}{The past two decades} have been characterized by
dramatic advances in the performance of organic semiconductor
devices, giving rise to the field known as organic electronics
\cite{A}. Light-emitting diodes \cite{B}, thin film transistors
\cite{C}, and photovoltaic cells \cite{D}, are examples of devices
being developed based on organic semiconductors. Critical to the
operation of all these devices is the process of charge injection
from a metal electrode into the organic semiconductor. The
efficiency of organic light emitting diodes, for example, is
directly related to the ability of the contacts to supply the
organic bulk with charge \cite{E}. Despite the great technological
importance of charge injection, the physics of this process
remains poorly understood. This may be ascribed to the fact that
transport in organic semiconductors is very different from that in
their inorganic counterparts. In the former materials, transport
takes place by hopping between localized electronic states, which
are distributed in energy due to spatially correlated  energetic
disorder with the standard deviation $\sigma \simeq 0.1$ eV
\cite{dip2,dpk}.

Recently, it was recognized that energetic disorder in organic
materials used in today's devices affects the injection efficiency
 \cite{conwell,1,2}. First, disorder was shown to increase
injection and, second, it was proposed as a major reason for the
unusually weak temperature dependence of the injected current
\cite{3,4}.

The injection properties of metal/organic interfaces depend on the
properties of a thin organic layer directly contacting with the
metal. It is well known that the structure of this interface layer
is typically different from the bulk structure of the organic
material. For this reason we may suspect that the variance
$\sigma^2_i$ of the energetic disorder at the interface differs
from the variance $\sigma^2_b$ of the disorder in the bulk of the
organic material. In literature \cite{conwell,1,2}, a calculation
of the effect of energetic disorder on the injection has been
carried out using bulk disorder parameters (basically, its
variance $\sigma^2_b$). To some extent this could be explained by
 lack of any detailed knowledge of the structure of the interface
layer. In addition,  experimental data of temperature dependence
of the  injected current seem to supports the idea that
$\sigma_i\approx\sigma_b$ \cite{3}. At the same time, it is well
known that frequently a surface dipole layer is formed directly at
the interface, providing an abrupt leap in carrier energy in the
range of $0.3-1$ eV \cite{ishii}. It is reasonable to assume that
such a layer has some degree of disorder and, thus, induces
additional energetic disorder in neighboring layers of organic
materials \cite{4}. The magnitude of this energetic disorder
should decay while going away from the interface, so $\sigma_i >
\sigma_b$, but this magnitude is completely unknown; in the
calculations carried out in Ref. \cite{4} very speculative
parameters have been used to estimate $\sigma_i$.

We are going to show that this problem of the relative magnitude
of $\sigma_i$ and $\sigma_b$ has an additional and quite
remarkable twist, because in organic devices sandwiched between
conducting electrodes the bulk disorder itself depends on the
proximity to the electrode.
 A well known fact is that a significant part of the total
energetic disorder in organic materials has an electrostatic
origin. For polar materials this is dipolar disorder
\cite{dip2,dpk,dip1}, while for non-polar materials it is
quadrupolar disorder \cite{q2}. Our major goal is to demonstrate
that the electrostatic disorder at the vicinity of metal/organic
interface differs from the bulk disorder far away from the
electrode.

Indeed, the electrostatic energetic disorder in organic materials
is directly proportional to the disorder in the spatial
distribution of electrostatic potential, generated by randomly
situated and oriented dipoles or quadrupoles. In organic layers
sandwiched between conducting electrodes this spatial distribution
must obey a boundary condition at the electrode surface: at this
surface the potential should be a constant. Thus, \textit{at the
electrode surface there is no energetic disorder at all},
irrespectively to how disordered is the material in the organic
layer. This means that the magnitude of the dipolar or quadrupolar
disorder \textit{increases while going away from the interface},
asymptotically reaching its bulk value. Here we assume that there
is no significant increase of the dipolar or quadrupolar disorder
directly at the interface (i.e., a disordered surface dipole layer
is absent). Now we are going to support this general idea with the
calculation of the variance of the dipolar disorder in the
vicinity of a conducting electrode.

Let us consider the simplest model of a rigid disordered polar
organic material where the randomly oriented (and orientationally
uncorrelated) point dipoles  occupy the sites of a simple cubic
lattice with the lattice scale $a$ \cite{dip1,dip2}. We consider
the vicinity of a
 metal electrode located at $z=0$, so the lattice occupies the
 half-space $z >0$  with the first lattice layer having
distance $a/2$ from the electrode plane. Charge carrier energy at
any site $m$ is the sum
\begin{equation}\label{Usum}
  U(\vec{r}_m)=e\sum_{n\neq m}\phi(\vec{r}_m,\vec{r}_n)
\end{equation}
where $\phi(\vec{r},\vec{r}_n)$ is the electrostatic potential,
generated by the dipole, located at the site $n$.  The variance of
the disorder is
\begin{equation}
\label{var} \sigma^2(\vec{r}_m)=\left<U_m^2\right>=e^2\sum_{n,l
\ne m} \left<\xi_n\xi_l\phi\left(\vec{r}_m,\vec{r}_n\right)
\phi\left(\vec{r}_m,\vec{r}_l\right)\right>,
\end{equation}
where the angular brackets denote the average over positions and
orientations of dipoles, and the variable $\xi_n$ equals to 1 if a
dipole is located at the site $n$ and 0 otherwise (note that
$\left<U_m\right>=0$). A spatial average gives
\begin{equation}
\label{pos}
\left<\xi_n\xi_l\right>=c\delta_{nl}+c^2\left(1-\delta_{nl}\right),
\end{equation}
where $c$ is the fraction of sites occupied by dipoles, and taking
into account that the average over dipole orientations gives
$\left<\phi\left(\vec{r}_m,\vec{r}_n\right)\right>=0$, we obtain
\begin{equation}
\label{var1} \sigma^2(\vec{r}_m)=e^2c\sum_{n \ne m}
\left<\phi^2\left(\vec{r}_m,\vec{r}_n\right)\right>.
\end{equation}
From this point the angular brackets denote the orientational
average only. In the case of an infinite medium without any
electrodes
\begin{equation}
\phi(\vec{r},\vec{r}_n)=\frac{\vec{p}_{n}\cdot \left( \vec{r}-\vec{r}_{n}%
\right) }{\varepsilon \left| \vec{r}-\vec{r}_{n}\right|^3 },
\label{phi_inf}
\end{equation}
where $\varepsilon$ is the dielectric constant of the medium and
$\vec{p}_n$ is the dipole moment. In the case of semi-infinite
medium bounded by an electrode, a boundary condition $\phi=0$ at
$z=0$ has to be imposed (we choose the arbitrary constant to be
zero). As a result, the source function $\phi(\vec{r},\vec{r}_n)$
includes a contribution from the image dipole
$\vec{p}_n^{\hskip2pt\textrm{i}}=(-p_{nx},-p_{ny},p_{nz})$ located
at $\vec{r}_n^{\hskip2pt\textrm{i}}=(x_{n},y_{n},-z_{n})$. an
average over dipole orientations gives
$\left<p_{ni}p_{nj}\right>=\frac{1}{3}p^2\delta_{ij}$ (with the
obvious modification for
$\left<p_{ni}p_{nj}^{\hskip2pt\textrm{i}}\right>$), and finally
\begin{eqnarray}
\label{sigma2(z)_direct} \sigma^2(z)=\frac{e^2 p^2
c}{3\varepsilon^2} \sum_{z_n>
0}\left[\frac{1}{\left|\vec{r}_n-\vec{z}\right|^4}+\frac{1}{\left|\vec{r}_n+\vec{z}\right|^4}-\right.\\
\left.2\frac{
r^2_n-z^2}{\left|\vec{r}_n-\vec{z}\right|^3\left|\vec{r}_n+\vec{z}\right|^3}\right],\nonumber
\end{eqnarray}
here the vector $\vec{z}=(0,0,z)$ and $\sigma$ does not depend on
$x$ and $y$. The lattice site with $\vec{r}_n = \vec{z}$ is
excluded from the sum (\ref{sigma2(z)_direct}). Note that
$\sigma(0)=0$, as it should be, because if the electrostatic
potential is a constant for $z=0$, then there is no electrostatic
disorder at this plane, no matter how many dipoles occupy the
half-space $z
> 0$. Far away from the electrode
\cite{dip2}
\begin{equation}
\sigma^2(\infty)=\sigma^2_b=\frac{e^2 p^2
c}{3\varepsilon^2}\sum_{r_n \ne 0} \frac{1}{r_n^4}\approx
5.51\frac{e^2 p^2 c}{\varepsilon^2 a^4}. \label{lattice}
\end{equation}

\begin{figure}
\begin{center}
\includegraphics[width=3in]{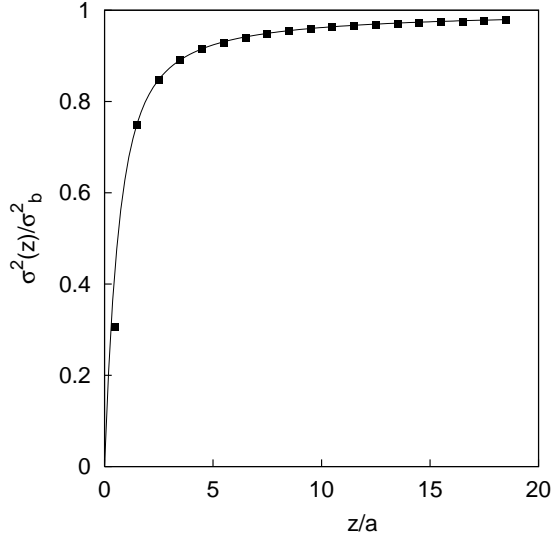}
\end{center}
\caption{Dependence of $\sigma^2(z)$ on $z$: the solid line
present the result of Eq. (\ref{sigma2(z)_3}) for $a_0=0.76a$,
while the squares correspond to the direct summation using Eq.
(\ref{sigma2(z)_direct}).} \label{fig1}
\end{figure}

We can perform an approximate analytical summation in Eq.
(\ref{sigma2(z)_direct}) if we provide the alternative expression
for the source function $\phi\left(\vec{r},\vec{r}_0\right)$, in
close analogy to the method, used in Ref. \cite{synthmet}. The
source function for the point dipole, located at $\vec{r}_0$, is
the solution of the Poisson equation
\begin{equation}
\label{Poisson_dipole} \Delta \phi=-\frac{4\pi
}{\varepsilon}\vec{p}\cdot \nabla_{\vec{r}_0}
\delta(\vec{r}-\vec{r}_0)
\end{equation}
and, hence, is proportional to the gradient of the Green function
$G\left(\vec{r},\vec{r}_0\right)$ of the Laplace operator  with
the zero boundary condition at $z=0$
\begin{equation}
\label{solution} \phi(\vec{r},\vec{r}_0)=-\frac{4\pi
}{\varepsilon}\vec{p}\cdot
\nabla_{\vec{r}_0}G\left(\vec{r},\vec{r}_0\right).
\end{equation}
Replacing summation with integration in Eq. (\ref{var1}) we have
\begin{equation}\label{sigma2(z)_2}
\sigma^2(z)\approx\frac{16\pi^2 e^2 p^2 c}{3\varepsilon^2 a^3}
\int_{z'>0} d\vec{r'}\left[\nabla_{\vec{r'}}
G(\vec{z},\vec{r'})\right]^2
\end{equation}
where the Green function has the form \cite{synthmet}
\begin{eqnarray}\label{G(r,r')}
G(\vec{r},\vec{r'})=\frac{1}{4\pi^2}\int d\vec{k}
e^{i\vec{k}(\vec{\rho}-\vec{\rho}\hskip2pt ')}G_k(z,z'),\\
 G_k(z,z')=-\frac{\sinh
kz_{-}}{k}\exp(-kz_{+}),\nonumber\\
z_{+}={\rm max}(z,z'),
 \hskip10pt z_{-}={\rm min}(z,z'),\nonumber
\end{eqnarray}
and $\vec{\rho}=(x,y)$ and $\vec{k}$ are two-dimensional
 vectors. Performing integration over
$\vec{\rho}\hskip2pt '$ in Eq. (\ref{sigma2(z)_2}) we obtain
\begin{widetext}
\begin{eqnarray}\label{sigma2(z)_3}
\sigma^2(z)\approx\frac{8\pi e^2 p^2 c}{3\varepsilon^2 a^3}
\int_0^\infty dkk \int_0^\infty dz'\left[ k^2
G_k^2(z,z')+\left(\frac{dG_k}{dz'}\right)^2\right]\approx\\
\frac{4\pi e^2 p^2 c}{3\varepsilon^2 a^3} \int_0^{1/a_0} dk
\left(1-e^{-2kz}\right)= \sigma^2_b
\left[1-\frac{a_0}{2z}\left(1-e^{-2z/a_0}\right)\right],
\hskip10pt \sigma^2_b= \frac{4\pi e^2 p^2 c}{3\varepsilon^2
a^3a_0}.\nonumber
\end{eqnarray}
\end{widetext}
Here a cut-off at $k\approx 1/a_0$ with $a_0\simeq a$ has been
introduced to remove the divergence at $k\rightarrow \infty$. This
cut-off is equivalent to the exclusion of the site with
$\vec{r}_n=\vec{z}$ in Eq. (\ref{sigma2(z)_direct}). We did not
introduce a similar short range cut-off in the integral over $z'$
in Eq. (\ref{sigma2(z)_3}) because this integral converges and the
possible correction does not change the result in a qualitative
way.
 To obtain an agreement
between the bulk $\sigma_b$ in Eq. (\ref{sigma2(z)_3}) and the
corresponding exact value for the lattice model in Eq.
(\ref{lattice}) we have to set $a_0\approx 0.76 a$ \cite{SPIE97}.
This choice of $a_0$ leads to a remarkably good agreement between
the approximate analytic expression (\ref{sigma2(z)_3}) and the
result of the direct summation according to Eq.
(\ref{sigma2(z)_direct}) in the whole range of distance from the
interface (see Fig. \ref{fig1}).

\begin{figure}
\begin{center}
\includegraphics[width=3in]{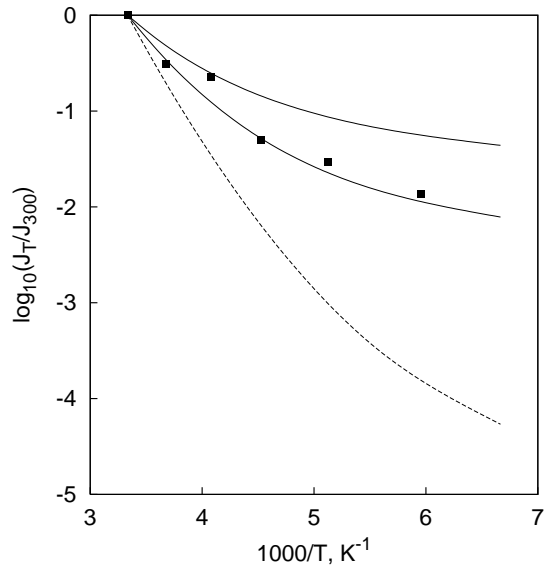}
\end{center}
\caption{Plot of the temperature dependence of the injected
current density $J_T$ (normalized by the corresponding density
$J_{300}$ for 300 K) for the injection of holes from the Ag
electrode into poly-dialkoxy-p-phenylene vinylene, calculated
using Eq. (\ref{arkh}) for $\sigma(z)=\textrm{const}$ (solid
lines) and $\sigma(z)$, calculate by Eq. (\ref{sigma2(z)_3})
(broken line), correspondingly. The squares indicate the
experimental points from Ref. \cite{3}. The upper solid line has
been calculated for $d=a/2$, and the lower one for $d=a$,
correspondingly.} \label{fig2}
\end{figure}

Note that for the transport sites situated within a distance of
$5-6$ lattice sites to the interface,  the amplitude of energetic
disorder is significantly decreased in comparison to its bulk
value. Yet this very thin layer is of crucial importance for
injection  in organic devices. We anticipate that the reduction of
$\sigma$ should lead to a stronger temperature dependence for the
injected current density in comparison with the
 treatment with $\sigma(z)=\textrm{const}$, provided in the paper by
Arkhipov \textit{et al}. \cite{1}. According to this model, the
injected current density is
\begin{widetext}
\begin{equation}
J \propto \int_d^\infty dz_0 \exp(-2\gamma
z_0)w_{\textrm{esc}}(z_0)\int_{-\infty}^\infty dU'
\textrm{Bol}(U')g\left[U_0(z_0)-U'\right], \hskip10pt
U_0(z_0)=\Delta-\frac{e^2}{4\varepsilon z_0}-eEz_0
 \label{arkh}
\end{equation}
\end{widetext}
where $\textrm{Bol}(U)=\exp(-U/kT)$ for $U > 0$ and
$\textrm{Bol}(U)=1$ otherwise, and $w_{\textrm{esc}}(z)$ is the
probability for a carrier to avoid the surface recombination
\begin{equation}
w_{\textrm{esc}}(z_0)=\frac{\int_d^{z_0}
dz\exp\left[-\frac{U_0(z)}{kT}\right]} {\int_d^{\infty}
dz\exp\left[-\frac{U_0(z)}{kT}\right]}. \label{wesc}
\end{equation}
Here $E$ is the applied electric field, $g(U)$ is the density of
states in the organic material (a Gaussian density of states is
usually assumed), $\gamma$ is the inverse localization radius,
$\Delta$ is the interface barrier, and $d$ is the minimal
distance, separating the electrode surface and the first layer of
the organic material. A natural generalization of the Eq.
(\ref{arkh}) to our case is straightforward: we have to let the
density of states $g$ depend on $z$ through the Eq.
(\ref{sigma2(z)_3}). We performed the calculation using parameters
provided in Ref. \cite{3} for the injection of holes from the Ag
electrode into poly-dialkoxy-p-phenylene vinylene: $\Delta=0.95$
eV, $\gamma=0.33$ \AA$^{-1}$, $E=5\times10^5$ V/cm,
$\sigma_b=0.11$ eV, and $d=12$ \AA $\hskip2pt$ (it was assumed
that $d=a$). Note that all these parameters were used in Ref.
\cite{3} for the comparison between the experimental data and Eq.
(\ref{arkh}) not as fitting parameters, but have been measured
independently. The result of the calculation is shown in Fig.
\ref{fig2}. A transformation of the curve occurs as anticipated:
because of the smaller $\sigma$ at the interface, the temperature
dependence of the  injected current becomes stronger and does not
agree with experimental data anymore. In fact, the agreement
between the experimental points and the curve, calculated using
Eq. (\ref{arkh}) for $\sigma(z)=\textrm{const}$, is not as perfect
as it appears to be, because we have to expect that $d < a$ is a
better choice for the minimal distance to the electrode ($a$ is
the intersite separation). For $d < a$ the curve, calculated by
Eq. (\ref{arkh}) for $\sigma(z)=\textrm{const}$, goes up (see Fig.
\ref{fig2}, the upper solid curve), and the agreement worsens.
Additionally, small distances to the electrode are very important
in the integral (\ref{arkh}), so the use of the macroscopic
$\varepsilon$ in Eq. (\ref{arkh}) is dubious. Again, the decrease
in $\varepsilon$ moves the curve for $\sigma(z)=\textrm{const}$ in
Fig. \ref{fig2} upwards.

All these reservations nothwithstanding, if we believe that the
model of Arkhipov \textit{et al}. \cite{1} is valid, then the
significant discrepancy between the lower broken line and the
experimental points in Fig. \ref{fig2} seems to be an indication
of the need to introduce an additional disorder (with $\sigma
\simeq 0.1$ eV)  at the interface. The possible source of this
additional disorder could be the surface dipole layer. From this
point of view, the experimental results, provided in Ref. \cite{3}
and connected to our consideration, in fact support the idea of a
disordered surface dipole layer: we clearly need additional
disorder at the interface to compensate the decrease in the
electrostatic disorder, provided by the bulk molecules.

Possible generalizations (taking into account the possible
additional spatial disorder at the interface, roughness of the
metal/organic interface) do not change our main conclusion that
the electrostatic part of the energetic disorder in organic
materials, provided by molecules in the bulk of the material, is
significantly suppressed at the interface.

If a disordered surface dipolar layer is indeed formed at the
interface, the picture suggested in this paper has to be modified.
In this case the magnitude of the total energetic disorder could
decrease with the increase of the distance to the electrode, or it
could still increase, depending on the relative amplitudes of the
bulk and surface contributions.

In conclusion, we have shown that the energetic disorder in
organic semiconductors decreases dramatically in the neighborhood
of a metal electrode. The ramification of this study is that
disorder parameters derived from bulk measurements do not describe
first few layers near the metal. As a result, simple models that
predict enhancement of charge injection in organic semiconductors
due to presence of disorder need to be reexamined.

 This work was supported by the ISTC grant 2207 and RFBR
grants 02-03-33052 and 03-03-33067. The research described in this
publication was made possible in part by Award No. RE2-2524-MO-03
of the U.S. Civilian Research \& Development Foundation for the
Independent States of the Former Soviet Union (CRDF).

\end{document}